# On Liquid Viscosity Effects on Droplet Splash and Receding Breakup on a Smooth Solid Surface at Atmospheric Pressure

Lei Yang,[†] Zhonghong Li,[†] Tao Yang,[‡] Yicheng Chi,[‡] and Peng Zhang[*,‡]

[†]College of Civil and Transportation Engineering, Shenzhen University, Shenzhen, 518060, China

[‡]Departmental of Mechanical Engineering, The Hong Kong Polytechnic University, Hung Hom, Kowloon, Hong Kong

**Abstract:** Experimental and modelling study is presented for the effect of a wide range of liquid viscosities on the droplet impact on a smooth solid surface at atmospheric pressure. A non-monotonic variation of threshold between droplet deposition and splash was observed experimentally. Specifically, the critical Weber number separating deposition from splashing decreases and then increases with increasing the Ohnesorge number. The observations of splash in low viscosity region and receding breakup in high viscosity region were analyzed qualitatively from the perspectives of Kelvin-Helmholtz instability and Rayleigh-Taylor instability, respectively. Based on instability analysis for the viscosity-induced nonmonotonicity, a new semi-empirical correlation of droplet splashing thresholds is proposed by fitting experimental results from previous and present data and shows better performance than previous correlation formulas.

[*] Corresponding author

Email: pengzhang.zhang@polyu.edu.hk.

Fax: (852)23654703, Tel: (852)27666664.

## 1. INTRODUCTION

The phenomenon of splashing during droplet impact on a solid surface is commonly encountered in many natural and industrial processes, such as raindrop, a dew drop dripping on soil, ink-jet printing, spray combustion, and various spraying processes in chemical, pharmaceutical and metallurgical engineering[1-6]. Various concomitant phenomena and physical aspects have been thoroughly summarized in a few excellent review papers[7-10], and the threshold of droplet splashing is the focus of the present study. It is well known that the splashing threshold depends on the physical properties of the droplet (including droplet size $D$, density $\rho$, surface tension $\sigma$, viscosity $\mu$, velocity $U$), those of the surface (including roughness[11, 12], wettability[13, 14], and stiffness[15, 16]), and the pressure and composition of surrounding gas[17-21]. A comprehensive parameterization of the threshold of droplet splashing for the above physical properties is often technically challenging and physically complex. Like many previous studies, the present one is concerned with droplet impact on a smooth solid surface in atmospheric pressure, so the Weber number, $We = \rho D U^2/\sigma$, and Ohnesorge number, $Oh = \mu/(\rho \sigma D)^{1/2}$, are the mere macroscopic variables used to quantify the complex droplets impact outcomes, such as spreading, rebound and splashing[2, 8, 10]. Nevertheless, the present study aims to clarify the effects of liquid viscosity that varies over wider ranges of $Oh$ and $We$.

Rioboo et al.[22] identified two distinct kinds of splashing (namely, prompt splash and corona splash) among six kinds of impact outcomes, such as deposition, prompt splash, corona splash, receding break-up, partial rebound, and complete rebound, for droplets impacting on a dry surface. In addition, they concluded that corona splashing is more likely to occur on a smooth surface, but the prompt splashing is prone to occur on a rougher surface. Mundo et al.[23] proposed that $K = OhRe^{5/4}$, where $Re = \rho UD/\mu$ is the Reynolds number, can be used to delineate the deposition and splashing regions, and incipient splashing occurs for $K > 57.7$. Vander Wal et al.[24] found the splash/non-splash boundary for a drop impact on a dry solid surface is well described by $\sqrt{Ca} = Oh\sqrt{Re} = 0.35$, where $Ca = \mu U/\sigma$ is the Capillary number. It is noted that Hao[12] recently found that surface roughness of micron scales has a nonmonotonic effect on corona splashing. Consequently, the present study gave special treatment to the surface to ensure negligible roughness.

Xu et al.[20] observed that gas compressibility is an essential factor resulting in splashing, and that decreasing the pressure of the surrounding gas may completely suppress the splashing. Subsequently, Xu et al.[19] aimed at establishing the relationship between substrate roughness and surrounding gas pressure in splashing dynamics, and they substantiated that prompt splashing is mainly related to surface roughness, while the instabilities due to the presence of the surrounding gas are essential factors causing corona splashing. Although there are different explanations about the mechanism of splashing[25-28], the perspective of hydrodynamic instabilities was found suitable for explaining the experimental observations of the present study.

Regarding the effects of liquid viscosity on droplet splashing, there are scattered discussions in the literature. Xu et al.[20] theoretically predicted an interesting but nonintuitive result: a more viscous liquid splashes more easily than a less viscous one. This result was subsequently validated by their own experiments on three different alcohols (methanol, ethanol and 2-propanol). It is noted that the kinematic viscosities of these alcohols vary about a factor of 3 (0.68-2.60 mm$^2$/s). Focusing on the threshold gas pressure at which splashing occurs, Xu[18] observed a non-monotonic variation of the pressure with liquid viscosity spanning more than one order of magnitude (0.68-18 mm$^2$/s). Specifically, the threshold gas pressure decreases and then increases with increasing the liquid viscosity. In the low-viscosity regime, the viscosity only affects the thickness of the expanding liquid film, and a larger viscosity causes a thicker film which is easier to destabilized due to the smaller surface tension force. In the high-viscosity regime, viscous drag is important and helps stabilize the spreading liquid film. This observation was also confirmed in the experimental study of Driscoll et al.[29], who suggested to examine the two viscosity regimes separately.

On the basis of above noteworthy studies, it is verified that liquid viscosity plays an important role in affecting the splashing threshold for droplet impacting on a smooth solid surface. However, the liquid viscosity that was considered in previous experimental studies varies either within an insufficiently large range or for a narrow range of impact velocity. Consequently, the range of Ohnesorge number and Weber number considered in those studies are inadequately wide to draw a confirmative conclusion about the viscosity effect in splashing threshold. In the present experiment, we investigated droplet splashing over sufficiently wide ranges of Weber number and Ohnesorge number with emphasis on obtaining a more complete understanding about the effects of the latter at different values of the former. Scaling correlations of t droplet splashing thresholds were proposed to compare with the experimental results.

## 2. EXPERIMENTAL SPECIFICATIONS

**2.1. Experimental Setup.** The experimental setup for droplet impacting on a smooth solid surface is sketched in Figure 1. A liquid droplet was generated from a syringe pump (LSP02-1B, Longer, China) with a stainless-steel needle of 360 μm outer diameter. Slowly pushing syringe pump with a flow rate of 10 μL/min to force liquid through the needle until the droplet is detached due to gravity. After triggering the photoelectric sensor, the process of droplet impacting on the solid surface was recorded by a high-speed camera (FASTCAM SA-X2, Photron, Japan) with a spatial resolution of 1024×672 and at a frame rate of 20,000 fps. The reproducible droplet was released from different heights, and its terminal velocity can be obtained by using the high-speed images at the moments before droplet impacting onto the surface. The experiments were carried out at atmospheric pressure and room temperature (20 ± 1°C).

**2.2. Characteristics of solid surface.** The present study aims at investigating liquid viscosity effects on droplet splashing on a solid surface. However, the characteristics of solid surface, which often introduce hidden variables into the problem, have a significant influence on droplet splashing. Besides the well-recognized surface characteristics such as roughness and wettability, surface stiffness also affect droplet splashing. According to Howland et al.'s study[16], solids with Young's moduli ≲ 100 kPa weaken droplet splashing. To reduce the surface stiffness effect on droplet splashing, we deliberately made a solid surface for the present study. The specific processing technology is as follow.

5The solid surface was made from Polydimethylsiloxane (PDMS, Sylgard 184, Dow Corning, Germany). The raw material was mixed with two kinds of components, monomer and cross-linker, at the mass ratio of 10 to 1. When the two components mixed thoroughly and was laid around 5-minute at room temperature, the mixture was poured into a stainless-steel mould placed on the top of silicon wafer and then degassed in a vacuum chamber for 30 minutes. After that, the mixture was roasted in an oven (ST-120B1, Espec, Japan) at 125°C for exactly 40 minutes. The thickness of the sample is 10 mm and the Young's modulus is 1.22 MPa which is much larger than 100 kPa. Therefore, the effect of surface stiffness in the experiment can be negligible. In addition, the surface used in the experiment is sufficiently smooth. The average surface roughness ($R_a$) of the sample is measured less than 50 nm by Atomic Force Microscopy (XE15, Park System, South Korea), so that the influence of surface roughness can be neglected in the problem.

**2.3. Characterization of droplets.** Nine water-glycerol solutions were made by mixing with distilled water at the different mass frictions of 0%, 18%, 30%, 45%, 60%, 68%, 75%, 80%, and 85%, respectively. The dynamic viscosity, $\mu$ varied from 1.005 mPa·s to 109 mPa·s, and the corresponding kinematic viscosity, $\nu$ varied from 1.0 mm$^2$/s to 91 mm$^2$/s. The diameter of droplets, $D$ varies in a range of 1.80~2.45mm. Other physical properties of these different liquids are shown in Table 1. It is seen that the viscosity can change significantly by about two orders of magnitude, while the changes of density and surface tension are within 21% and 7%, respectively. By using the solutions, we were able to study the droplet splashing by varying $Oh$ while fixing $We$ or vice versa.

3. **RESULTS AND DISCUSSION**

    **3.1. Phenomenological observations.** The present experiment investigated comprehensively the effects of liquid viscosity (characterized by the Ohnesorge number, $Oh$) and the impact velocity (characterized by the Weber number, $We$). All the testing cases covering a range of $Oh$ from 0.002 to 0.3 and a range of $We$ from 20 to 550 are summarized in Figure S1-S6 of the Supporting Information. Several representative cases of droplet impact are shown in Figure 2, in which a droplet of 30% glycerol solution freely falls from different heights (100 mm, 200 mm, 400 mm and 800 mm) to impact on the

6smooth solid surface. We can clearly see that the splashing appears in the spreading process of liquid film from falling at 800 mm height, while there was no droplet disintegration in either the spreading process or the receding process for other cases, as a lower height results in a lower impact velocity and a smaller Weber number. Increasing the impact velocity promotes droplet splashing has been well known in the previous studies. We shall focus on the results with varying $Oh$ and at fixed Weber numbers.

As shown in Figure 3, no splashing occurred all long in the five cases with different $Oh$ ($5.98 \times 10^{-3}$, $1.12 \times 10^{-2}$, $2.56 \times 10^{-2}$, $8.40 \times 10^{-2}$, and $2.59 \times 10^{-1}$) when $We$ is fixed around 143. With increasing $We$ to be around 246, an interesting nonmonotonic phenomenon emerges, as shown in Figure 4, that no splashing occurred at relatively small viscosities, splashing then appeared in the spreading process for the droplet of 60% glycerol solution ($Oh = 2.56 \times 10^{-2}$), and no splashing occurred at higher viscosities. As $We$ is increased to around 359, the phenomenon was more conceivable that the splashing appeared in lower-viscosity cases but disappeared for the higher-viscosity case of 85% glycerol solution, as shown in Figure 5. It should be noted that the droplets of the first three cases (30%-60% glycerol solution) splashed during the liquid film spread, but the breakup of high $Oh$ liquids (75%-85% glycerol solution) happened as the film rim was retracting. As $We$ is increased to about 453, as shown in Figure 6 (Multimedia view) taken from video 1-5, splashing occurred for all the cases. Similarly, high $Oh$ liquids only broke up within the retracting period.

**3.2. Non-monotonic dependence of droplet splashing on liquid viscosity.** Experimental results of splashing and non-splashing are presented as a regime nomogram in the $Oh$-$We$ parameter space, as shown in Figure 7. It is seen that, within the ranges of $We$ concerned, there are three distinct consequences of varying $Oh$. In the relatively low $We$ range, no splashing occurs regardless of the value of $Oh$. This is consistent with the previous observation that splashing favours a sufficiently large impact velocity. For the same reason, splashing occurs for all the $Oh$ in the relatively high $We$ range. The remaining but interesting phenomenon to be interpreted is that, in the intermediate range of $We$, droplet splashing varies non-monotonically with $Oh$. Specifically, in the smaller $Oh$ range (0.003~0.01), the threshold $We$ decreases with $Oh$, indicating that increasing viscosity promotes splashing; in the larger $Oh$ range (0.04~0.3), the threshold $We$ increases with $Oh$, indicating that increasing viscosity suppressed splashing.

The promotion of droplet splashing by viscosity has been observed and interpreted in the previous studies. By following Xu et al.'s analysis[18] based on the Kelvin-Helmholtz (K-H for short hereinafter) instability, we proposed a scaling interpretation to the present observation as follows. First, droplet splashing is assumed to occur when the thickness $d$ of the liquid film formed during droplet spreading becomes comparable with the wavelength of the most unstable K-H wave[30], namely

$$\frac{d}{k_{KH}^{-1}} = O(1) \tag{1}$$

According to the previous studies[18, 31, 32], the film thickness governed by viscous boundary layer is

$$d \sim \sqrt{\mu t/\rho} \tag{2}$$

$k_{KH} \sim \Sigma_G/\sigma$ is the wavenumber of the most unstable K-H wave, where $\Sigma_G$ is the restraining stress of the gas film on the spreading liquid. It is proportioned to the spreading velocity $V_e$, but it does not depend on the liquid surface tension and viscosity[18]. Consequently, we have

$$\alpha = \frac{d}{k_{KH}^{-1}} \sim \Sigma_G \sqrt{\frac{\mu t}{\rho}}/\sigma \propto \sqrt{\mu}/\sigma \tag{3}$$

Therefore, droplet splashing becomes easier with the increase of $\mu$, and its threshold $We$ is assumed to vary with viscosity in the same way. It is seen in Figure 7 that this correlation agrees very well with the experimental results. Physically, the agreement substantiates the hypothesis discussed in the introduction that, in the low-viscosity regime, the liquid film is stabilized by surface tension, and the thinner film with decreasing the viscosity makes it more stable, indicating that a larger impact Weber number is needed to make splashing to happen.

For droplet splashing in the higher-$Oh$ range, we note that droplet splashing occurs when the droplet has reached the maximum spreading and started to retract back, as shown in Figure 8. The retraction of the liquid film is characterized by an acceleration motion[33] and therefore subject to the Rayleigh-Taylor (R-T for short hereinafter) instability. It is noted that the secondary droplet formation during the liquid film retraction was classified as a distinct phenomenon called as "receding breakup"[22, 34]. In the present context, we fully recognized that this receding breakup is phenomenologically similar to droplet splashing but governed by different mechanism.

By adopting the perspective of hydrodynamic instability, we again assume that droplet receding break-up occurs when the film thickness is comparable with the most unstable R-T wave and have

$$\frac{d_m}{k_{RT}^{-1}} = O(1) \tag{4}$$

where $k_{RT}$ is the wavenumber of the most unstable R-T wave. Same with Eq. (2), the film thickness is formulated as

$$d_m \sim \sqrt{\mu t_m/\rho} \tag{5}$$

where $t_m$ is the time for the largest spreading time. According to the theory of R-T instability[35],

$$k_{RT} \propto \sqrt{a} \sim \sqrt{u_c/\tau_v} \tag{6}$$

where the acceleration $a$ of droplet receding can be estimated by using the Taylor-Culick velocity[36-39], $u_c = \sqrt{2\sigma/\rho d_m}$ and its viscous relaxation time $\tau_v = \mu d_m/2\sigma$. Although R-T instability is derived based on inviscid theory and is possibly not a nicety in high-viscosity situation, it is competent to reflect the leading effect of viscosity on the droplet breakup. Consequently, the R-T wave has a correlation with viscosity-relevant terms

$$k_{RT} \propto \sqrt{u_c/\tau_v} \propto \frac{(1/d_m)^{1/4}}{\sqrt{\mu d_m}} \tag{7}$$

Substituting Eq. (5) and Eq. (7) into Eq. (4), we have

$$\alpha = \frac{d_m}{k_{RT}^{-1}} \propto d_m^{\frac{1}{4}} \mu^{-\frac{1}{2}} \propto \mu^{-\frac{3}{8}} \tag{8}$$

Therefore, Eq. (8) suggests that droplet splashing becomes harder with the increase $\mu$, and its threshold $We$ is assumed to vary with the viscosity in the same way, which agrees well with the experimental results in Figure. 7.

It should be noted that the above physical interpretations for the influence of liquid viscosity on the threshold *We* are mostly phenomenological. The derived relations, Eq. (3) and Eq. (8), cannot be used as scaling laws for predictions, because they do not reflect the correct flow similarity. The physically sound scaling laws should be derived based on the analysis on both viscous liquid film and the beneath gas film and will be considered in our future study. In addition, there are secondary effects of the liquid viscosity to be considered. First, taking into account for liquid viscosity in the K-H and R-T instability analysis would result in corrections to the wavenumbers of the most unstable waves. Second, the viscous drag can be important in the high-viscosity regime and helps to stabilize the film. Direct experimental evidence for the instabilities merits future studies.

**3.3. Empirical scaling laws for the deposition and splash threshold.** Previous studies[20, 23, 24, 30, 40, 41] have proposed many splashing correlations in terms of different pairs of parameters among $Oh$, $We$, and $Re$. Mundo et al.'s $K = OhRe^{5/4}$, where $K = 57.5$, indicates a monotonic dependence ($\sim \mu^{-1/4}$) on liquid viscosity. Consequently, it cannot be used to represent the nonmonotonic viscosity effects on droplet splashing, shown in Section 3.2. For droplet splash under different ambient pressures, Stevens[42] fitted two distinct regimes where the threshold pressure has a inflexion with varying impacting velocity. Because only low-viscosity fluids were considered in Stevens' experiment, $We$ and $Re$ in his fitting formula are always positively correlated. In Riboux and Gordillo's theoretical model[41] of drop splash, the splashing criteria can be divided into low-$Oh$ and high-$Oh$ regimes, with scaling relations being $t_{e,crit} \propto Re^{1/2}$ and $t_{e,crit} \propto We^{-2/3}$, respectively, where $t_{e,crit}$ is the dimensionless ejection time. This implicitly verifies the nonmonotonic viscosity effects and is consistent with the present experimental observations. Palacios et al.[40] proposed a splashing threshold as

$$We = 5.8Re^{1/2} + 4.01 \times 10^7 Re^{-1.97} \qquad (9)$$

Since their experiments were limited to $Re > 400$, Eq. (9) shows a deviation for splashing droplets with larger liquid viscosities.

Recognizing that an empirical correlation is often desirable for practical purposes, we propose a new semi-empirical correlation for deposition and splashing threshold as

$$We = 5Re^{1/2} + 10^4 Re^{-3/4} \qquad (10)$$

By comparing Eq. (10) with several sets of experimental data and previous criteria from the literature[23, 40, 41, 43], as shown in Figure 9, we can make the following observations. First, the present formula can produce overall good agreement with the previous experimental data over the entire $Re$ regime, while other formulas show discrepancies either in the lower $Re$ regime or higher $Re$ regime. Second, Palacios et al.'s formula, $5.75 = WeRe^{-1/2}$, agrees well for low viscosity droplets ($Oh < 0.01$). From Eq. (3), we can obtain the same exponents of $We$ and $Re$ as $\alpha \propto \sqrt{\mu}/\sigma \propto WeRe^{-1/2}$. This again confirms that the present physical interpretation based on the K-H instability may capture the essential physics of the splash of low viscosity droplets. Palacios et al. also proposed a threshold correlation Eq. (9), which is very similar to our formula Eq. (10), but it does not agree with Mundo et al.'s high viscous data and present receding breakup cases. Third, Mundo et al.'s formula, $57.7 = OhRe^{5/4}$, works for high viscosity droplets ($Oh > 0.04$), which is paralleled with the Ohnesorge line[23], $132.3 = OhRe^{5/4}$. The Ohnesorge line describes the decay of a liquid film, where the R-T instability is used to account for the receding breakup of high viscosity droplets as we discussed in Section 3.2. Fourth, it is noted that the splashing of mercury droplets in Schmidt and Knauss[43], used by Mundo et al. for supporting the $57.7 = OhRe^{5/4}$, are clearly nongregarious with other liquids. This suggests that there may be other physics, for example liquid volatility, affecting the mercury droplet impact. Fifth, it should be noted that the threshold splashing data from different experiments have some moderate discrepancies in the transition region of $Oh = 0.01 \sim 0.04$, which might be caused by their different experimental conditions.

It should be emphasized that the present Eq. (10) merely is a practically useful semi-empirical formula accounting for the influence of liquid viscosity on droplet impact at atmospheric pressure. It is by no means neither a rigorous scaling theory nor a universal correlation for the droplet splash[20, 42] under different ambient conditions. Undoubtedly, the influence of the ambient gas including its pressure and components cannot manifest in terms of $Oh - We - Re$ regimes, and it is beyond the scope of the present study and will be considered in our further work.

## 4. CONCLUSIONS

An experimental study of water-glycerol droplets impacting on a solid smooth surface at atmospheric pressure is performed in the range of $Oh = 0.002 \sim 0.3$ and $We = 20 \sim 550$, with particular emphasis on clarifying the effects of viscosity on the splashing process. The most important observation in the experiment is the non-monotonic dependence of droplet splashing on the liquid viscosity. The parametric characterization of the phenomenon is that, for intermediate $We$, the droplet splashing appears in the intermediate $Oh$ but is absence at relatively small and large $Oh$. To interpret the non-monotonic phenomenon, we adopted a perspective based on instabilities of the liquid film. The splashing occurring at small $Oh$ is controlled by the K-H instability, and the derived scaling relation Eq. (3) accords with the experimental observation that the threshold $We$ decreases with $Oh$. The receding breakup occurring at large $Oh$ is controlled by the R-T instability, and the derived scaling relation Eq. (8) accords with the experimental observation that the threshold $We$ increases with $Oh$. Besides, these arguments are supported by the previous scaling laws for deposition/splashing thresholds[23, 40]. Based on experimental observations and the instability analysis, a new semi-empirical splashing correlation Eq. (10) is proposed by fitting experimental data in a wide viscosity range. The underlying physics of droplet splashing, particularly the effect of ambient gas and pressure, based on interfacial instabilities merits future investigation and validation.

## SUPPLEMENTARY MATERIAL

See supplementary material on images and videos for droplets splashing with different glycerol solution impacting on a solid surface.

## AUTHOR INFORMATION

**Corresponding Author**

*Email: pengzhang.zhang@polyu.edu.hk

**ORCID**

Peng Zhang: 0000-0002-1806-4200

**Notes**

The authors declare no competing financial interest.

**ACKNOWLEDGMENTS**

The work at Shenzhen University was supported by National Natural Science Foundation of China (No. 11102116), and the work at the Hong Kong Polytechnic University was supported by the Hong Kong Research Grants Council/General Research Fund (PolyU 152651/16E) and partly by Central Research Grant (G-SB1Q).

**DATA AVAILABILITY**

The data that support the findings of this study are available from the corresponding author upon reasonable request.

Table 1. Physical properties of water-glycerol solutions at 1 atm and 20±1°C.

| Fluid | Glycerol (wt %) | Density ($10^3$ kg/m$^3$) | Surface tension (mN/m) | Viscosity (mPa·s) |
|---|---|---|---|---|
| Mixture 1 | 0 | 0.998 | 72.1 | 1.005 |
| Mixture 2 | 18 | 1.041 | 72.0 | 1.655 |
| Mixture 3 | 30 | 1.056 | 71.7 | 2.50 |
| Mixture 4 | 45 | 1.104 | 69.8 | 4.66 |
| Mixture 5 | 60 | 1.150 | 68.8 | 10.8 |
| Mixture 6 | 68 | 1.168 | 68.5 | 19.1 |
| Mixture 7 | 75 | 1.179 | 68.2 | 35.5 |
| Mixture 8 | 80 | 1.196 | 67.8 | 60.1 |
| Mixture 9 | 85 | 1.210 | 67.4 | 109 |

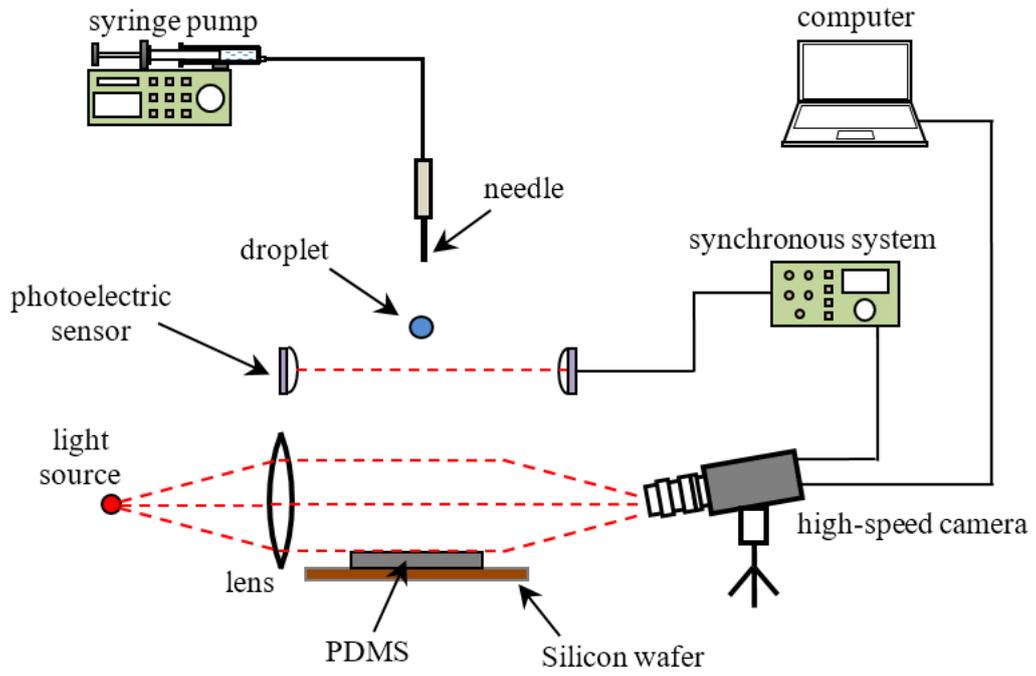

Figure 1. Sketch of experimental setup for reproducible droplet impact on a smooth solid surface.

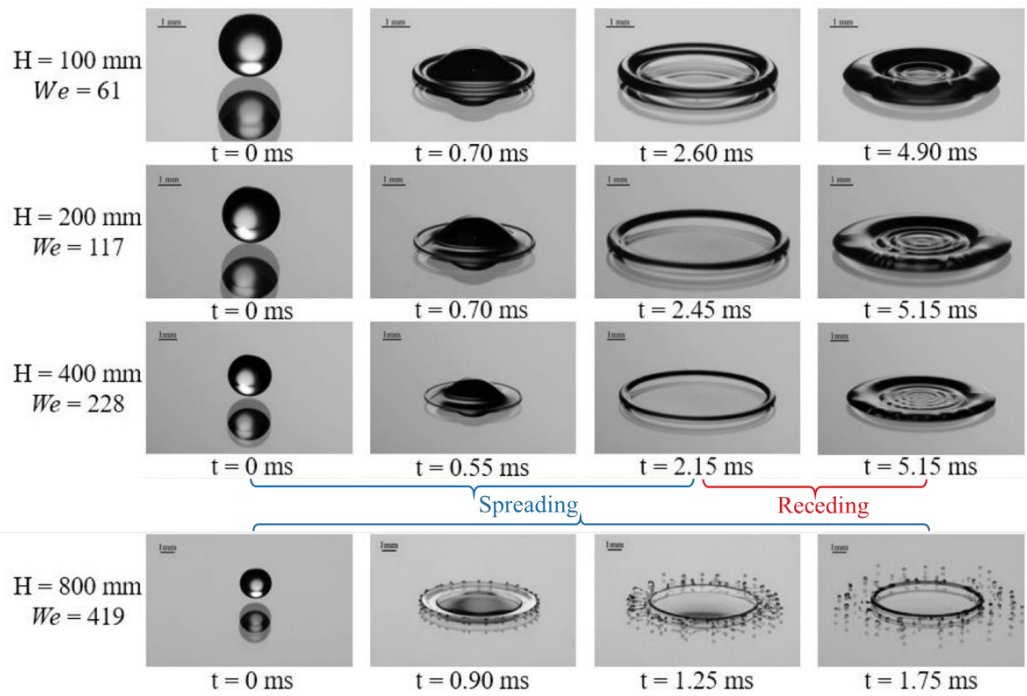

Figure 2. High-speed photographs of a droplet falling from different heights impacting on a solid surface, the droplet splash occurred during the spreading process of liquid film at $We = 419$ (30% glycerol solution, $D = 2.25$ mm, $Oh = 6.06 \times 10^{-3}$).

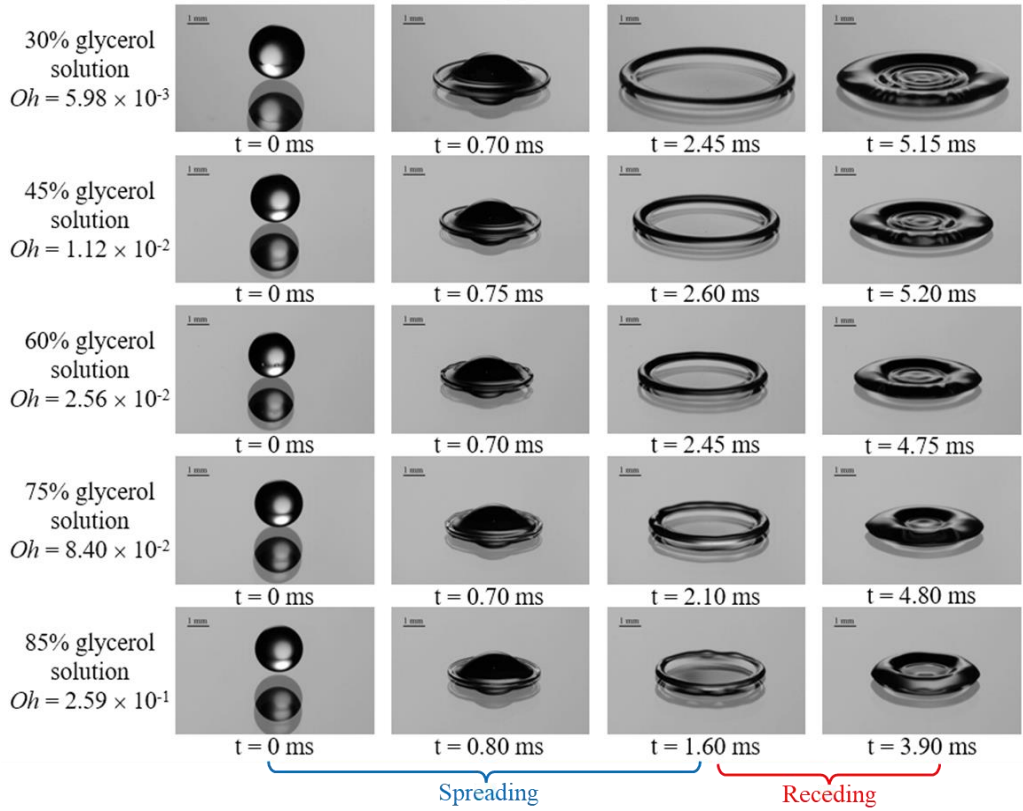

Figure 3. Images of droplets with different glycerol solutions impacting on a solid surface at $We = 143 \pm 26$.

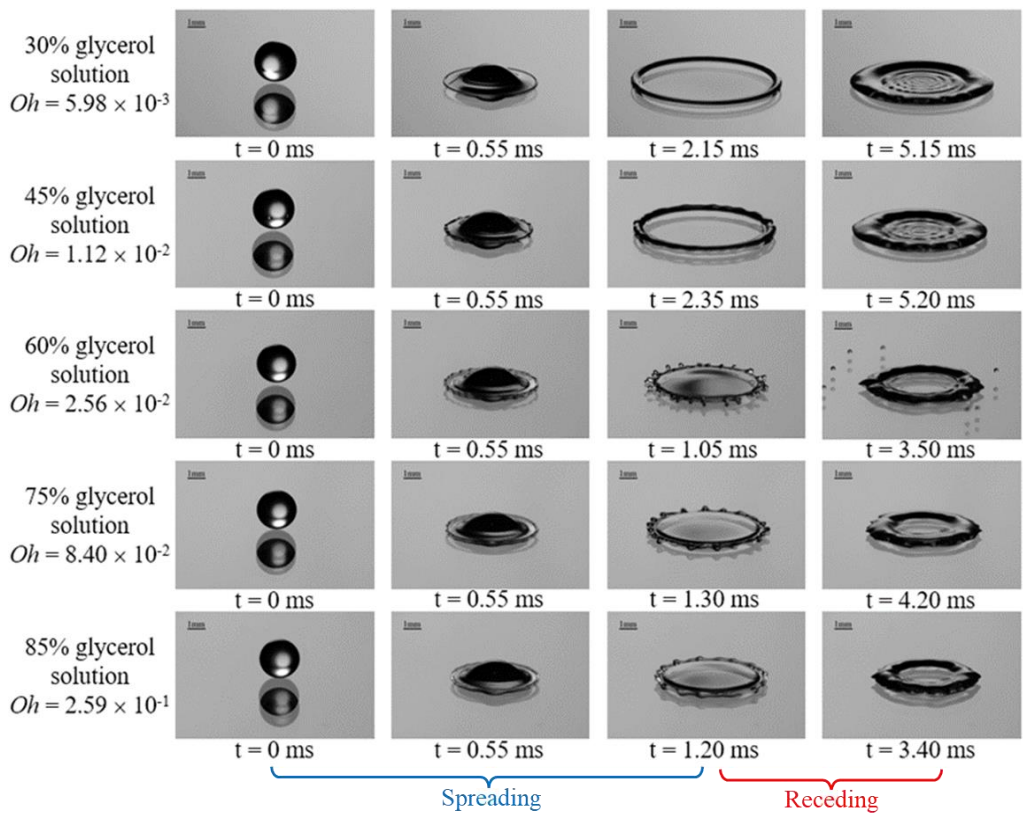

Figure 4. Images of droplets with different glycerol solutions impacting on a solid surface $We = 246 \pm 12$.

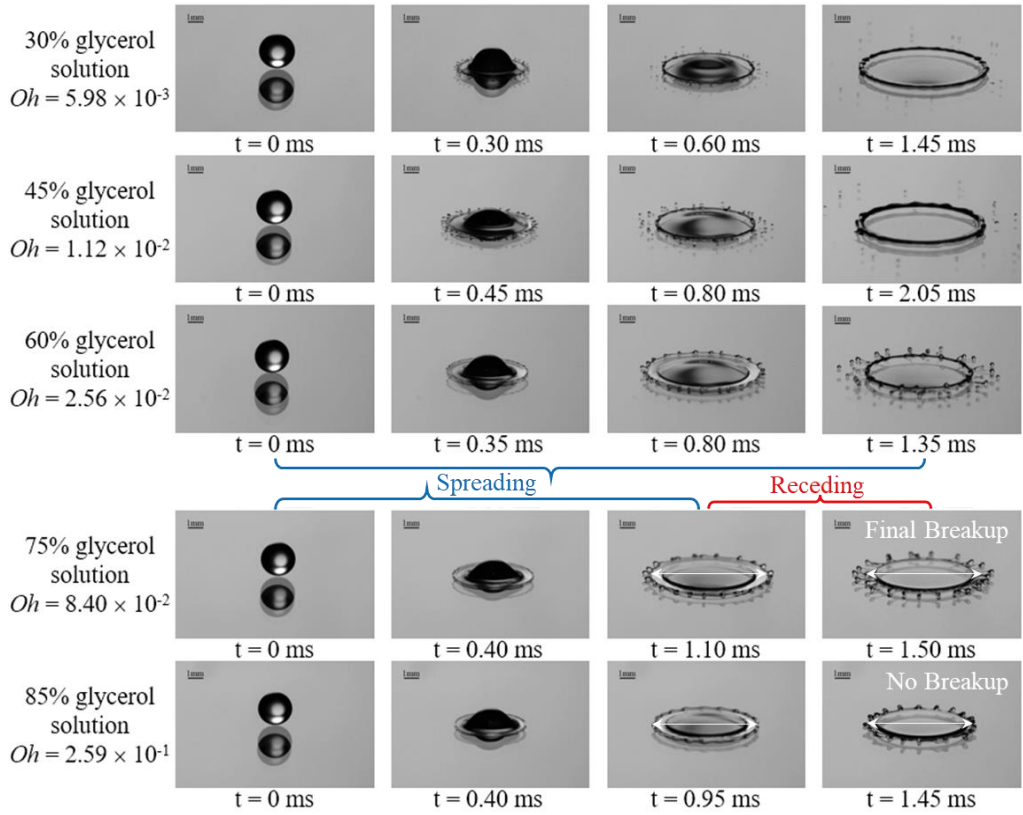

Figure 5. Images of droplets with different glycerol solutions impacting on a solid surface $We = 359 \pm 25$. The white arrows in each cases have same length.

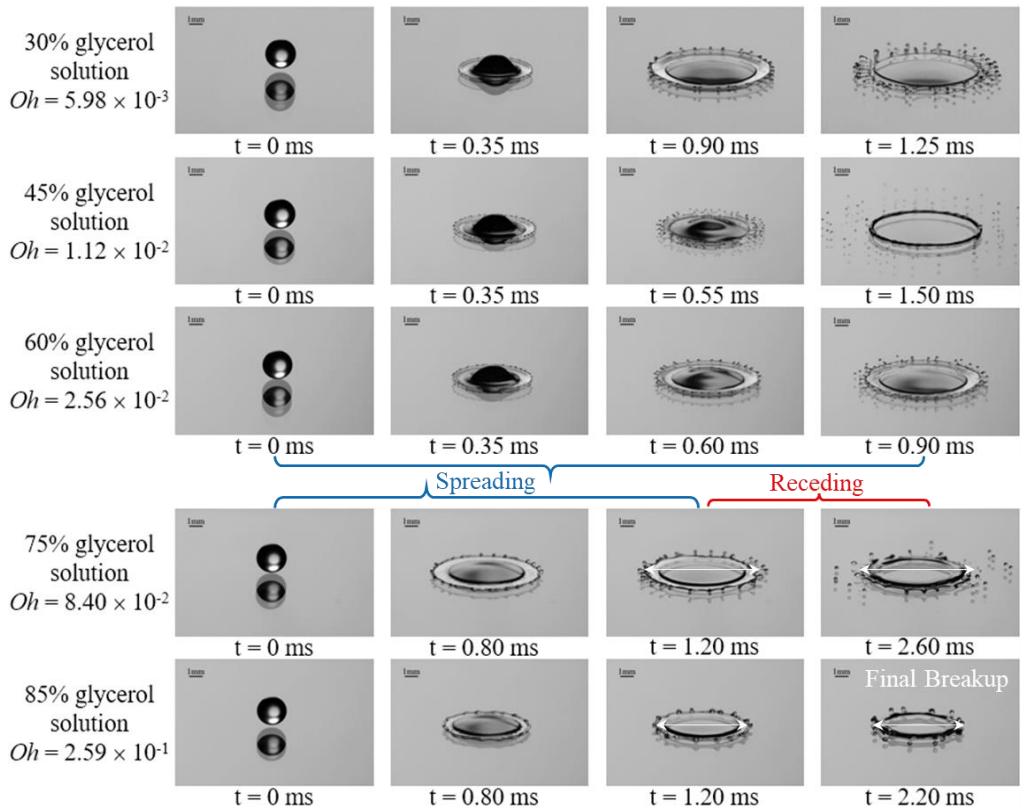

Figure 6. Images of droplets with different glycerol solutions impacting on a solid surface $We = 453 \pm 31$. The white arrows in each cases have same length. (Multimedia view)

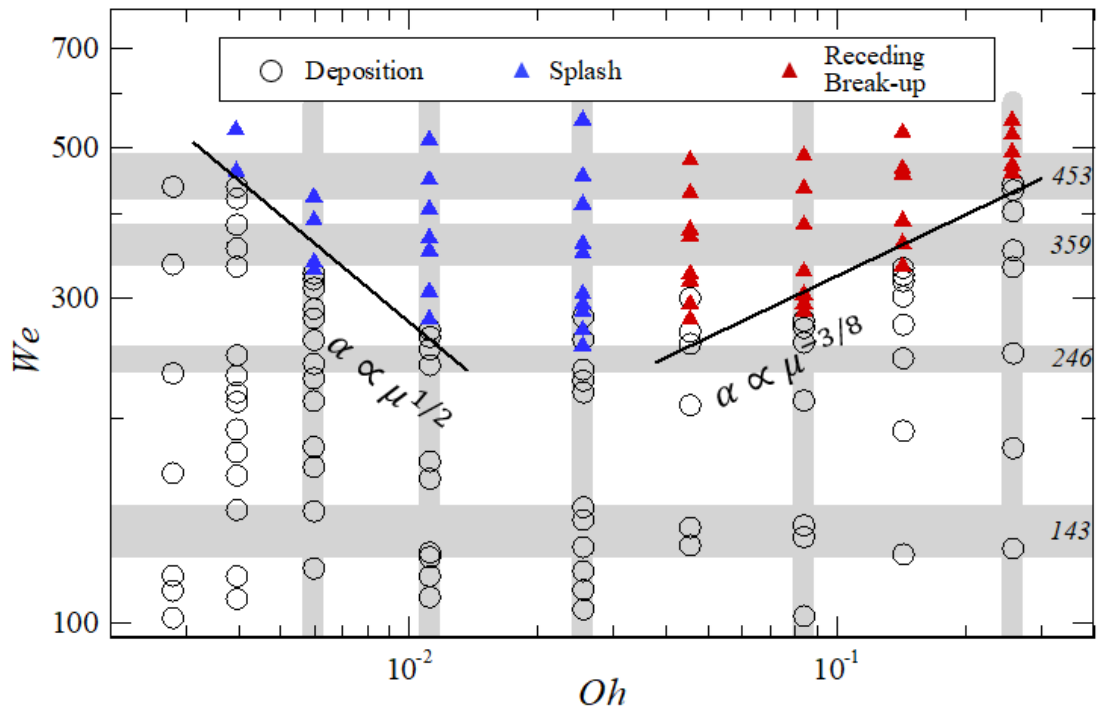

Figure 7. Splash and receding break-up regime nomogram in the $Oh - We$ subspace of $We = 100 \sim 550$ and $Oh = 0.002 \sim 0.3$, where the Eq. (3) holds the smaller $Oh$ range (0.003 ~ 0.01) and the Eq. (8) holds the larger $Oh$ range (0.04 ~ 0.3). The cases in Figure 3-6. corresponds the cross points in grey areas.

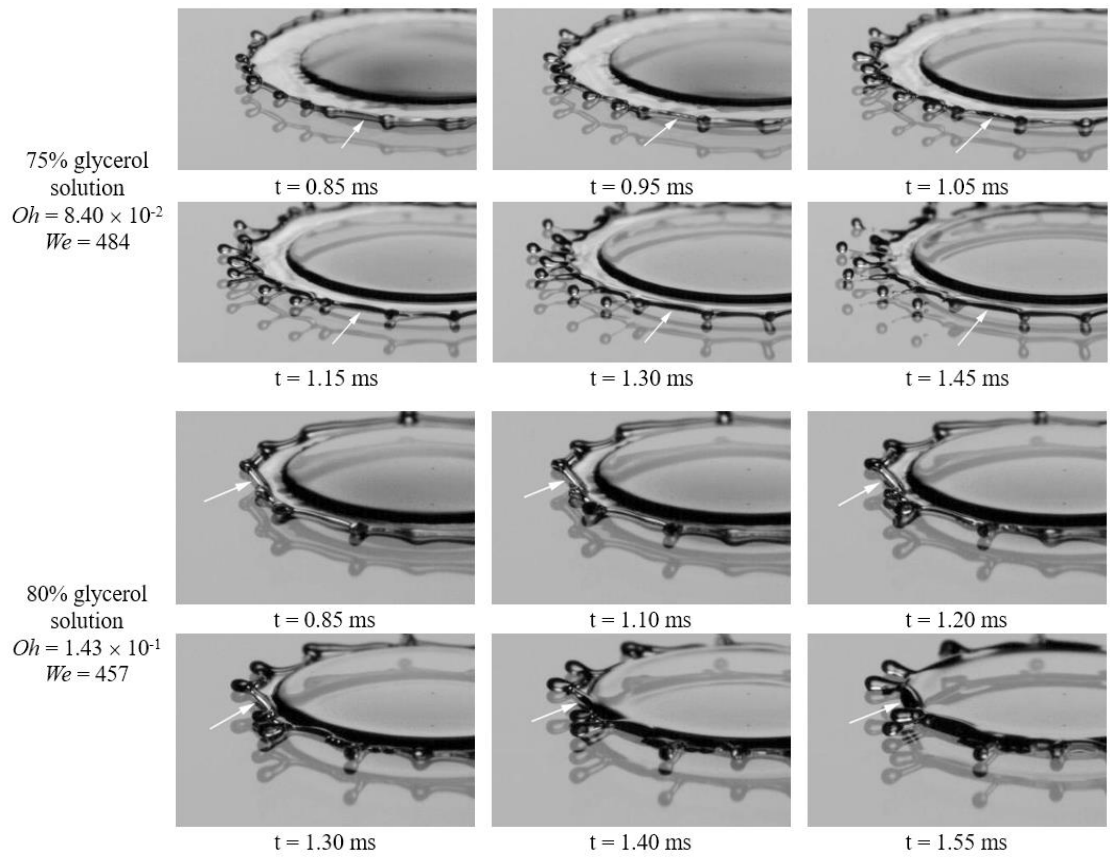

Figure 8. Images of fingering break-up from film rim (as shwon by arrows) during droplet receding process. (a) 75% glycerol solution, $Oh$ = 8.40×10$^{-2}$, and $We$ = 484, (b) 80% glycerol solution, $Oh$ = 1.43×10$^{-1}$, and $We$ = 457.

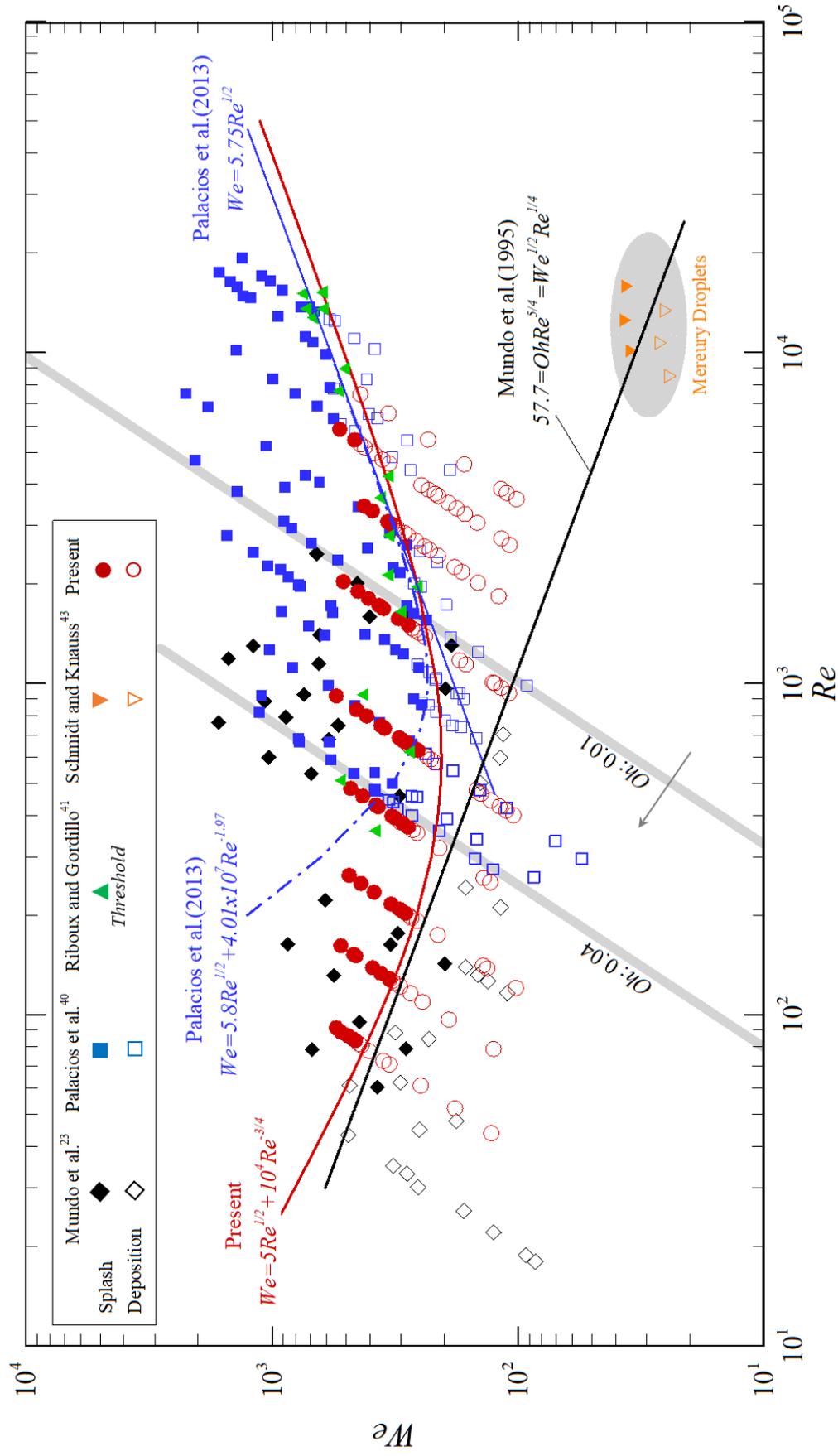

Figure 9. Comparison in the *We-Re* plane between a few threshold correlations with experimental data from Mundo et al.[23], Palacios et al.[40], Riboux and Gordillo[41], Schmidt and Knauss[43] and present splash and receding breakup cases.